\documentclass{iau}
\usepackage{graphicx}

\title[~~Quasars as tracers of cosmic flows] %% give here short title %%
{Quasars as tracers of cosmic flows}

\author[J. Modzelewska, B. Czerny, M. Bilicki, K. Hryniewicz et al.]   %% give here short author list %%
{J. Modzelewska$^1$,
B. Czerny$^1$, M. Bilicki$^2$, K. Hryniewicz$^{1,3}$,\\
  M. Krupa$^4$, F. Petrogalli$^1$, W. Pych$^1$,  A. Kurcz$^4$ \and A. Udalski$^5$}

\affiliation{$^1$Nicolaus Copernicus Astronomical Center, Bartycka 18, 00-716 Warsaw, Poland \\ email: {\tt jmodzel@camk.edu.pl}, {\tt bcz@camk.edu.pl}\\[\affilskip]
$^2$Astrophysics, Cosmology and Gravity Centre, Department of Astronomy,\\ University of Cape Town, Rondebosch, South Africa \\[\affilskip]
$^3$ISDC Data Centre for Astrophysics, Observatoire de Geneve,\\ Universite de Geneve, Chemin d'Ecogia 16, 1290 Versoix, Switzerland\\[\affilskip]
$^4$Astronomical Observatory of the Jagiellonian University, Orla 171, 30-244 Cracow, Poland\\[\affilskip]
$^5$Warsaw University Observatory, Al. Ujazdowskie 4, 00-478 Warszawa, Poland\\}

\pubyear{2014}
\volume{308}
\jname{The Zeldovich Universe: Genesis and Growth of the Cosmic Web}
\editors{R. van de Weygaert, S. Shandarin, E. Saar \& J. Einasto, eds.}

\begin{document}

\maketitle

\begin{abstract}
Quasars, as the most luminous persistent sources in the Universe, have broad applications for cosmological studies. In particular, they can be employed to directly measure the expansion history of the Universe, similarly to SNe Ia. The advantage of quasars is that they are numerous, cover a broad range of redshifts, up to $z = 7$, and do not show significant evolution of metallicity with redshift. The idea is based on the relation between the time delay of an emission line and the continuum, and the absolute monochromatic luminosity of a quasar. For intermediate redshift quasars, the suitable line is Mg II. Between December 2012 and March 2014, we performed five spectroscopic observations of the QSO CTS C30.10 ($z = 0.900$) using the South African Large Telesope (SALT), supplemented with photometric monitoring, with the aim of determining the variability of the line shape, changes in the total line intensity and in the continuum. We show that the method is very promising. 
\keywords{accretion disks - black hole physics, emission line, quasar - individual: CTS C30.10}
\end{abstract}

\begin{figure}[b]
% \vspace*{-2.0 cm}
\begin{center}
\begin{minipage}[htb]{.48\textwidth}
\centering
 \includegraphics[width=2.6in]{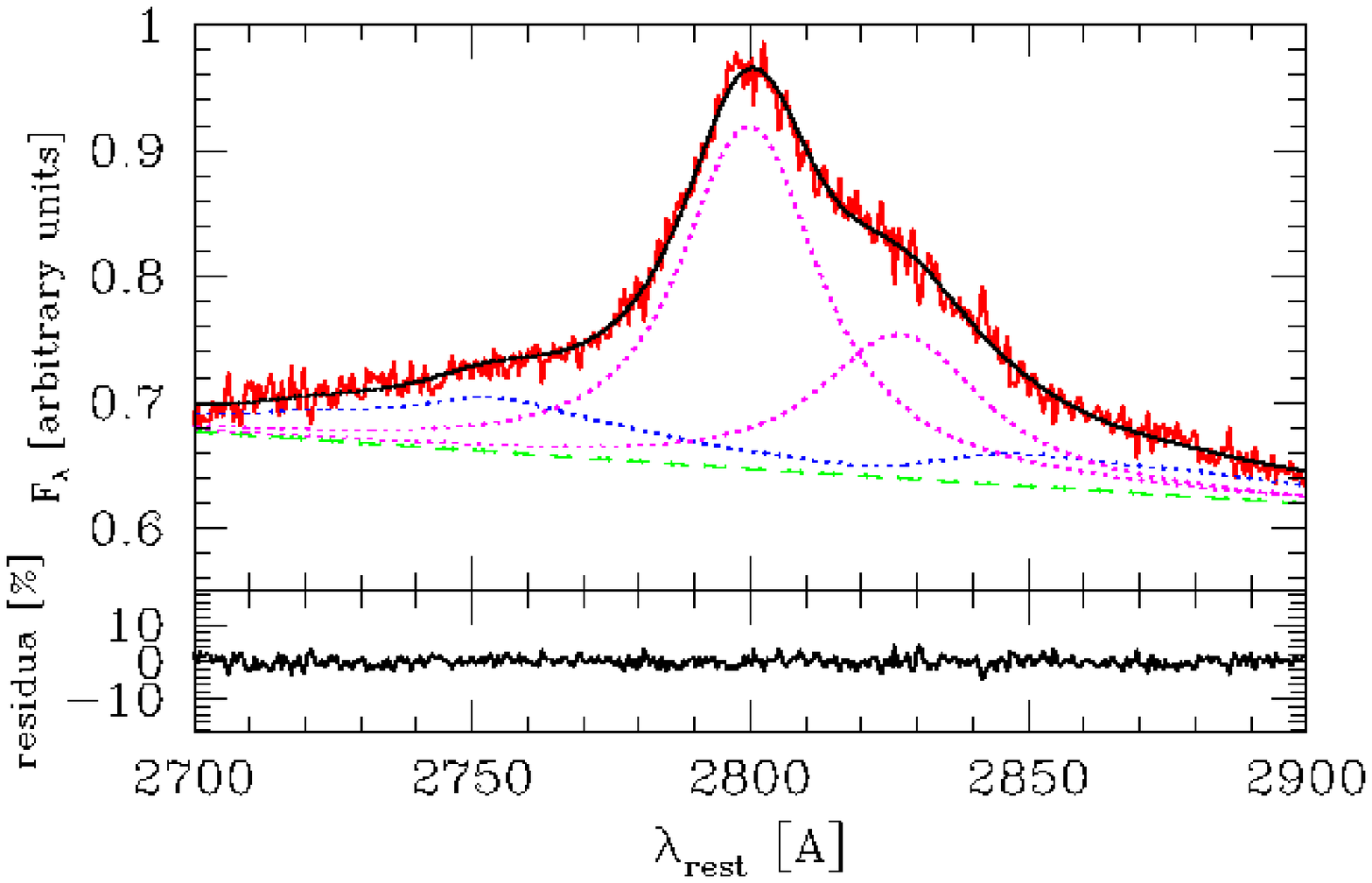} 
% \vspace*{-1.0 cm}
 \caption{Best fit and residual for the 1st observation of the two kinematic components in emission of the Mg II line (dotted lines) and FeII theoretical templates  [Bruhweiler \& Verner (2008), d12-m20-20-5]; continuous lines show the model and the data, dashed lines give the underlying power law.}
   \label{fig1}
\end{minipage}
\begin{minipage}[htb]{.48\textwidth}
\centering
\includegraphics[width=2.6in]{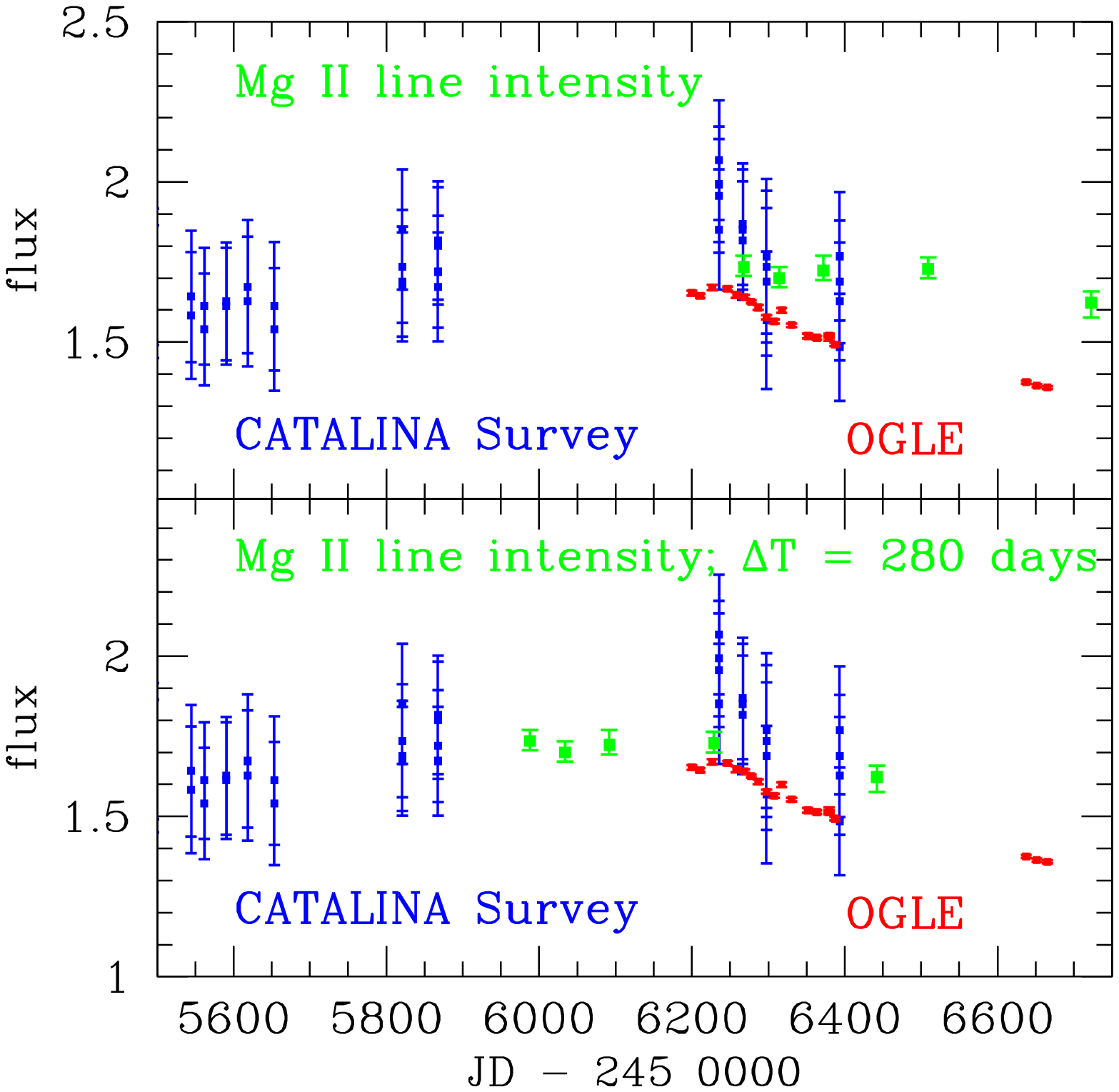} 
% \vspace*{-1.0 cm}
 \caption{Time evolution of the V-band flux and the MgII line intensity as measured (upper panel), and after a shift by 280 days corresponding to a plausible time delay (lower panel).}
   \label{fig2}
\end{minipage}
\end{center}
\end{figure}

\firstsection % if your document starts with a section,
\section{Introduction}
Quasars represent the high luminosity tail of active galactic nuclei (AGN). Among their multiple applications are probing the intergalactic medium (Borodoi et al. 2014) or providing information on massive black hole growth (Kelly et al. 2011), but they have been also proposed as promising tracers of the expansion of the Universe. The latter two aspects mostly rely on the presence of the Broad Emission Lines in quasar spectra. The BLR (Broad Line Region) is unresolved but the spectral variability allows to measure the size of the BLR from the time delay between the lines and of the continuum. For sources at redshifts $0.4 < z < 1.5$ the suitable line for such study is Mg II, monitored in the optical range. This delay is then used to determine the absolute quasar luminosity (see Czerny et al. 2013) employing the idea of dust origin of the BLR (Czerny \& Hryniewicz 2011).

\section{Results}
Using the South African Large Telescope (SALT), we obtained five spectra of the QSO CTS C30.10, taken in a period of 15 months. All the spectra were analysed separately, in a relatively narrow spectral range of 2700 -- 2900 \AA~ in the rest frame. We used 16 different FeII pseudo-continuum templates and we fit the spectra with the continuum and the Mg II line at the same time. The Mg II line in this source had to be modelled by two separate kinematic components, meaning that CTS C30.10 is a type B source. We considered two components in emission with a double Lorentzian line shape, which provided the best fit. Using photometry from the Optical Gravitational Lensing Experiment (OGLE), we were able to calibrate the spectra properly and to obtain the calibrated line and continuum luminosity. The time dependence of the SALT Mg II flux, and OGLE and CATALINA Survey continuum luminosity are shown in Fig. 2 (Modzelewska et al. 2014). The monitoring of the distant quasar for 15 months has not allowed yet for any firm conclusion on the time delay between the continuum and the Mg II line; however, we can try to make some preliminary estimates based on the fact that the continuum had a clear maximum just at the beginning of our monitoring campaign.

\section{Summary}
Reverberation studies of quasars can be used as new cosmology probes of the expansion of the Universe. The understanding of the formation of the BLR in AGN, and in particular of the properties of the Mg II line, is also important in a much broader context. The measurement important for cosmological applications is the time delay, and this can be determined well in type B sources. Our monitoring has been too short so far to allow for a detection, but the variability pattern of the line and the continuum seems to suggest a delay of about 300 days.

\end{document}